\documentclass{PoS}
\usepackage{amsmath}
\usepackage{booktabs}
\usepackage{braket}
\usepackage{nicefrac}
\usepackage{siunitx}
\usepackage{subcaption}
\usepackage{url}
\DeclareCaptionSubType*[alph]{table}
\DeclareCaptionSubType*[alph]{figure}
\title{Kaon-kaon scattering at maximal isospin from $N_f=2+1+1$ twisted
        mass lattice QCD}

\ShortTitle{Kaon-kaon scattering at maximal isospin from $N_f=2+1+1$
            twisted mass lattice QCD}

\author{\speaker{C. Helmes}, C. Jost, B. Knippschild, B. Kostrzewa, L. Liu, C. Urbach, M. Werner%
                 \\
                 University of Bonn\\
                 E-mail: \email{helmes@hiskp.uni-bonn.de}}
                                 
\abstract{
    We present results for the interaction of two kaons at maximal
    isospin. The calculation is based on 2+1+1 flavour gauge
    configurations generated by the ETM Collaboration (ETMC) featuring
    pion masses ranging from about 230 MeV to 450 MeV at three
    values of the lattice spacing. The elastic scattering length
    $a_0^{I=1}$ is calculated at several values of the bare
    strange quark and light quark masses. We find $M_K a_0
    =-0.397(11)(_{-8}^{+0})$ as the result of a chiral and continuum
    extrapolation to the physical point. This number is compared to other
    lattice results.}

\FullConference{34th annual International Symposium on Lattice Field Theory\\
		24-30 July 2016\\
		University of Southampton, UK}

\begin{document}
\section{Introduction}
We investigated the scattering of 2 $K^+$-mesons in the maximum isospin channel
of $I_3=1$. Due to the fact of $SU(3)$-symmetry breaking by the, compared to the
light quark, heavy strange quark, $\chi$-PT often suffers from sizeable
corrections. Lattice QCD offers a non-perturbative way to access this
interaction and helps in understanding the dynamics of the underlying strong
interaction at low energies. We focus on the elastic scattering length of this
system which has not been determined
experimentally. The analysis of the $K^+$-$K^+$-system proceeds very similarly to the one of $\pi$-$\pi$-scattering at $I_3=2$
detailed in ref.~\cite{Helmes:2015gla}. Our study is the first to take into
account three values of the lattice spacing.

\section{Elastic Scattering in L\"{u}scher's formalism}
The scattering length $a_0$ for the scattering of two mesons is defined by the low energy limit
\begin{equation} \lim_{q \rightarrow 0} q\cot\delta_0(q) =
  -\frac{1}{a_0}\,,
\end{equation}
where $q$ is the momentum transfer in the center of mass frame and $\delta_0$
the phaseshift of the outgoing wave function.
In a series of papers,~\cite{Luscher:1985dn}-~\cite{Luscher:1986pf}, L\"{u}scher related the scattering length to the
energy shift $\delta E$ of a system of two
particles~\cite{Luscher:1986pf} in a finite volume. This energy shift is the
deviation of the total energy of the interacting two particle system, $E_{KK}$, from its
expected value in the absence of the interaction between the two particles,
$2 E_K$.
For the case of two $K^+$-mesons at maximal isospin in the center of mass frame it reads:
\begin{equation}
\label{lusch_de}
\begin{split}
  \delta E_{KK}^{I=1} =E_{KK}-2E_{K}= &- \frac{4\pi a_0}{M_K L^3}
        \left[ 1 + c_1\frac{a_0}{L} + c_2\left(\frac{a_0}{L}\right)^2 \right]
        + O(L^{-6})\,,\\
        &c_1 = -2.9837297, \quad c_2 = 6.375183\,,
\end{split}
\end{equation}
with the spatial lattice extent $L$ and the kaon mass $M_K$, where the kaons are
at rest.

\section{Lattice Action and Operators}
We work with Wilson twisted mass Lattice QCD (tmLQCD) at maximal
twist introduced in ref.~\cite{Frezzotti:2000nk}.
This guarantees automatic $\mathcal{O}(a)$ improvement of all physical
quantities of interest as shown in ref.~\cite{Frezzotti:2003ni}. 
The gauge configurations have 2+1+1 dynamical quark flavours and were generated by the
ETMC. In total 3 different lattice spacings in the range from
\SIrange{0.619}{0.885}{\femto\metre} and a pion mass range from
\SIrange{230}{450}{\mega\electronvolt}
build a sound basis for continuum and chiral
extrapolations.
\begin{table}[t!]
 \centering
 {\small
 \begin{tabular*}{.9\textwidth}{@{\extracolsep{\fill}}lcccccc}
  \hline\hline
  ensemble & $\beta$ & $a\mu_\ell$ & $a\mu_\sigma$ & $a\mu_\delta$ &
  $(L/a)^3\times T/a$ & $N_\mathrm{conf}$  \\ 
  \hline\hline
  $A30.32$   & $1.90$ & $0.0030$ & $0.150$  & $0.190$  &
  $32^3\times64$ & $280$  \\
  $A40.24$   & $1.90$ & $0.0040$ & $0.150$  & $0.190$  &
  $24^3\times48$ & $404$  \\
  $A40.32$   & $1.90$ & $0.0040$ & $0.150$  & $0.190$  &
  $32^3\times64$ & $250$  \\
  $A60.24$   & $1.90$ & $0.0060$ & $0.150$  & $0.190$  &
  $24^3\times48$ & $314$  \\
  $A80.24$   & $1.90$ & $0.0080$ & $0.150$  & $0.190$  &
  $24^3\times48$ & $306$  \\
  $A100.24$  & $1.90$ & $0.0100$ & $0.150$  & $0.190$  &
  $24^3\times48$ & $312$  \\
  \hline
  $B35.32$   & $1.95$ & $0.0035$ & $0.135$  & $0.170$  &
  $32^3\times64$ & $250$ \\
  $B55.32$   & $1.95$ & $0.0055$ & $0.135$  & $0.170$  &
  $32^3\times64$ & $311$ \\
  $B85.24$   & $1.95$ & $0.0085$ & $0.135$  & $0.170$  &
  $32^3\times64$ & $296$ \\
  \hline
  $D30.48$ & $2.10$ & $0.0030$ & $0.120$ & $0.1385$ &
  $48^3\times96$ & $369$ \\
  $D45.32sc$ & $2.10$ & $0.0045$ & $0.0937$ & $0.1077$ &
  $32^3\times64$ & $301$ \\
  \hline\hline
 \end{tabular*}
 }
 \caption{The gauge ensembles used in this study. The labelling of
   the ensembles follows the notation in
   Ref.~\cite{Baron:2010bv}. In addition to the relevant input
   parameters we give the lattice volume and the number of evaluated
   configurations, $N_\mathrm{conf}$.}
 \label{tab:setup}
\end{table}
In this analysis we use twisted mass light valence quarks at maximal twist and
Osterwalder-Seiler (OS) strange valence quarks on a Wilson twisted
mass fermion-sea. This mixed action approach enables tuning of the valence
strange quark mass to its physical value without having to regenerate gauge
configurations. The cost of this is introducing small unitarity
breaking effects. The Dirac operator for the light sector (used as valence and
sea Dirac operator) reads:
\begin{equation}
  D_\ell = D_W + m_0 + i \mu_\ell \gamma_5\tau^3\, ,
  \label{eq:D_tm_val}
\end{equation}
with the Wilson Dirac operator $D_W$ and the Pauli matrices $\tau^i\,
,\,\,i=1,2,3$ acting in flavour space. The parameter $\mu_\ell$ denotes the twisted
mass $\pm\mu_\ell$ for the spinor $\chi_\ell$ on which it acts ($\chi_\ell =(u,
d)^T$). The spinors $\chi_\ell$ ($\bar{\chi}_\ell$) are connected to their physical counterparts, $\psi_\ell$ ($\bar{\psi}_\ell$), via a chiral rotation,
\begin{equation}
  \bar{\psi}_\ell = \bar{\chi}_\ell \exp\left(i\gamma_5\tau_3\frac{\omega}{2}\right)
  ,\quad \psi_\ell = \exp\left(i\gamma_5\tau_3\frac{\omega}{2}\right) \chi_\ell\,,
\end{equation}
around the twist angle $\omega$.
Working at maximal twist means $\omega=\pi/2$.
For the OS strange valence quarks we use the operator:
\begin{equation}
  D_s = D_W + m_0 + i \mu_s \gamma_5\,.
  \label{eq:D_os_val}
\end{equation}
Further details of the formulation of the OS action can be found
in ref.~\cite{Frezzotti:2004wz}.
The gauge configurations have been generated with the Iwasaki gauge action
of ref.~\cite{Iwasaki:1985we}.
The ensembles cover 11 values
of $a\mu_l$ at 3 different lattice spacings $a$.
We calculated strange quark propagators for 3 different strange quark
mass parameters $a\mu_s$ per lattice spacing. Tab.~\ref{tab:mus} states the corresponding values
of $a\mu_s$ for every $\beta$.
The one and two particle operators, in the physical basis denoted by $K(t)$ and
$\mathcal{O}_{KK}(t)$,
\begin{equation*}
  K(t) = \sum_{\vec{x}} \bar{\psi}_s(\vec{x},t)\gamma_5\psi_u(\vec{x},t)\,, \quad
  \mathcal{O}_{KK}(t)=K(t)K(t)\,,
\end{equation*}
respectively, lead to the correlation functions of the single kaon $C_K(t)$ and the two kaon
system $C_{KK}(t)$. They are defined as follows:
\begin{equation*}
    C_K(t) = \braket{K(t)K^{\dagger}(0)}\,, \quad
    C_{KK}(t) =\braket{\mathcal{O}_{KK}(t) \mathcal{O}^{\dagger}_{KK}(0)}\,.
\end{equation*}
We extract the ground state energy of each correlation function at times large
enough such that excited states have decayed sufficiently.
Because we work
at zero total momentum this approach yields the lattice kaon mass $M_K$
from $C_K$ and the total energy $E_{KK}$ of the $K^+$-$K^+$-system from $C_{KK}$.
Unfortunately, the spectral decomposition of $C_{KK}$ gets distorted by
terms constant in Euclidean time, preventing the naive extraction of
$E_{KK}$, as detailed in ref.~\cite{Feng:2009ij}.
Following ref.~\cite{Feng:2009ij} we use a ratio of correlation functions
\begin{align}
  \label{eq:ratio}
              R(t+a/2)=\frac{C_{KK}(t) - C_{KK}(t+a)}
              {C_{K}^2(t) - C_{K}^2(t+a)}\,,
\end{align}
which can be shown to behave, for large Euclidean times $t$, like
\begin{align}
  \label{eq:ratio_fit}
        R(t+a/2)\propto A_R\left( \cosh\left( \delta E_{KK} t'  \right) +
        \sinh\left( \delta E_{KK} t' \right)\coth\left(2M_{K}t'\right) \right)\,,
        \quad t'=t+\frac{a}{2}-\frac{T}{2}.
\end{align}
Here $T$ is the total time extent of the lattice.
$\delta E_{KK}$ is obtained by fitting eq.~\ref{eq:ratio_fit} to the lattice data of eq.~\ref{eq:ratio}.
We determine $a_0$ by inserting $M_K$, $L$ and $\delta E_{KK}$
into eq.~\ref{lusch_de} and solving for $a_0$.
\begin{table}
 \centering
 {\small
 \begin{tabular*}{.7\textwidth}{@{\extracolsep{\fill}}lrrr}
  \hline\hline
  $\beta$ & $1.90$ & $1.95$ & $2.10$ \\
  \hline\hline
  $a\mu_s$ & 0.0185 & 0.0160 & 0.013\\
           & 0.0225 & 0.0186 & 0.015\\
           & 0.0246 & 0.0210 & 0.018\\
  \hline\hline
  $a\,\, [\si{\femto\metre}]$ & 0.0885(36) & 0.0815(30) & 0.0619(18)  \\
  \hline\hline
 \end{tabular*}
}
 \caption{Values of the bare strange quark mass $a\mu_s$ used for the
   three $\beta$-values and values of the lattice spacing $a$}
 \label{tab:mus}
\end{table}
We apply quark field smearing in a Laplacian-Heaviside (LapH) manner
to our quark fields, as proposed in ref.~\cite{Peardon:2009gh}, and calculate all-to-all
propagators. To reduce the computational costs we combine the
LapH method with a stochastic approach (sLapH) as suggested in
ref.~\cite{Morningstar:2011ka}, resulting in 5 (3) random
vectors per
light (strange) quark perambulator used in the computation.
Diluting the random vectors further reduces the stochastic noise.
A more detailed description of the sLapH method can be found in reference~\cite{Helmes:2015gla} and references
therein.

\section{Analysis Strategies}
The bare quark mass parameters $a\mu_l$ and $a\mu_s$ are related to their
physical counterparts via the renormalisation constant $Z_P$: $m_f = \mu_s/Z_P$.
To work at the physical strange quark mass value we need to specify how to set
$m_s$ in our calculation. Setting the strange quark mass in different ways
serves as a consistency check.
\paragraph{Method A} consists of fixing $a\mu_s$ to the value that
reproduces the physical value of the difference
\begin{equation}
\Delta^2 = r_0^2(M_K^2-M_{\pi}^2/2)
\end{equation}
on each ensemble. The lattice scale is set using the Sommer Parameter 
$r_0 =
\SI{0.474(14)}{\femto\metre}$ and the values for the lattice spacing determined
in ref.~\cite{Carrasco:2014cwa}.
Tab.~\ref{tab:mus} states the values employed for each lattice spacing.
\paragraph{Method B} uses the experimental squared kaon mass to fix the strange
quark mass via a fit to data for the squared kaon
mass as a function of the light and strange quark masses
\begin{equation}
  \label{eq:fix_ms_methB_alt}
  (r_0 M_K)^2 =
  \bar{P}_0(r_0m_l+r_0m_s)\left[1+\bar{P}_1r_0m_l+\bar{P}_2\left(\frac{a}{r_0}\right)^2\right]K_{M_K^2}^{FSE}\,,
\end{equation}
which includes lattice artifacts of $\mathcal{O}(a^2)$~\cite{Carrasco:2014cwa}.
After the fit the continuum values of $(r_0M_K)^2$ and $r_0m_l$
from ref.~\cite{Carrasco:2014cwa} are used to determine the continuum
value of $r_0m_s$. 

With the values of $M_K a_0$ interpolated to the strange quark mass of Method A
or B we extrapolate the data for $M_K a_0$ in $m_l$ and $a$ to
the point of $m_l^{\mathrm{phys}}$ determined in
ref.~\cite{Carrasco:2014cwa} and the continuum. In leading order chiral
perturbation theory $M_K a_0$ depends linearly on $m_l$. Because
of automatic $\mathcal{O}(a)$-improvement we only have to consider
discretisation effects of $\mathcal{O}(a^2)$. In the continuum and
the limit $\mu_l\rightarrow 0$, we have a residual value of $M_Ka_0$, stemming from the
non-zero strange quark mass value, giving rise to the parameter
$P_2$ in eq.~\ref{eq:cont_ext}. Therefore, we fit the
function
\begin{equation}
  \label{eq:cont_ext}
  M_Ka_0 = P_0(r_0M_{\pi})^2 + P_1\left(\frac{a}{r_0}\right)^2 + P_2
\end{equation}
to our data at the physical strange quark mass.

\section{Results}
\begin{figure}
\vspace{-0.5cm}
  \begin{subfigure}{.49\textwidth}
    \flushleft
    \includegraphics[width=\textwidth]{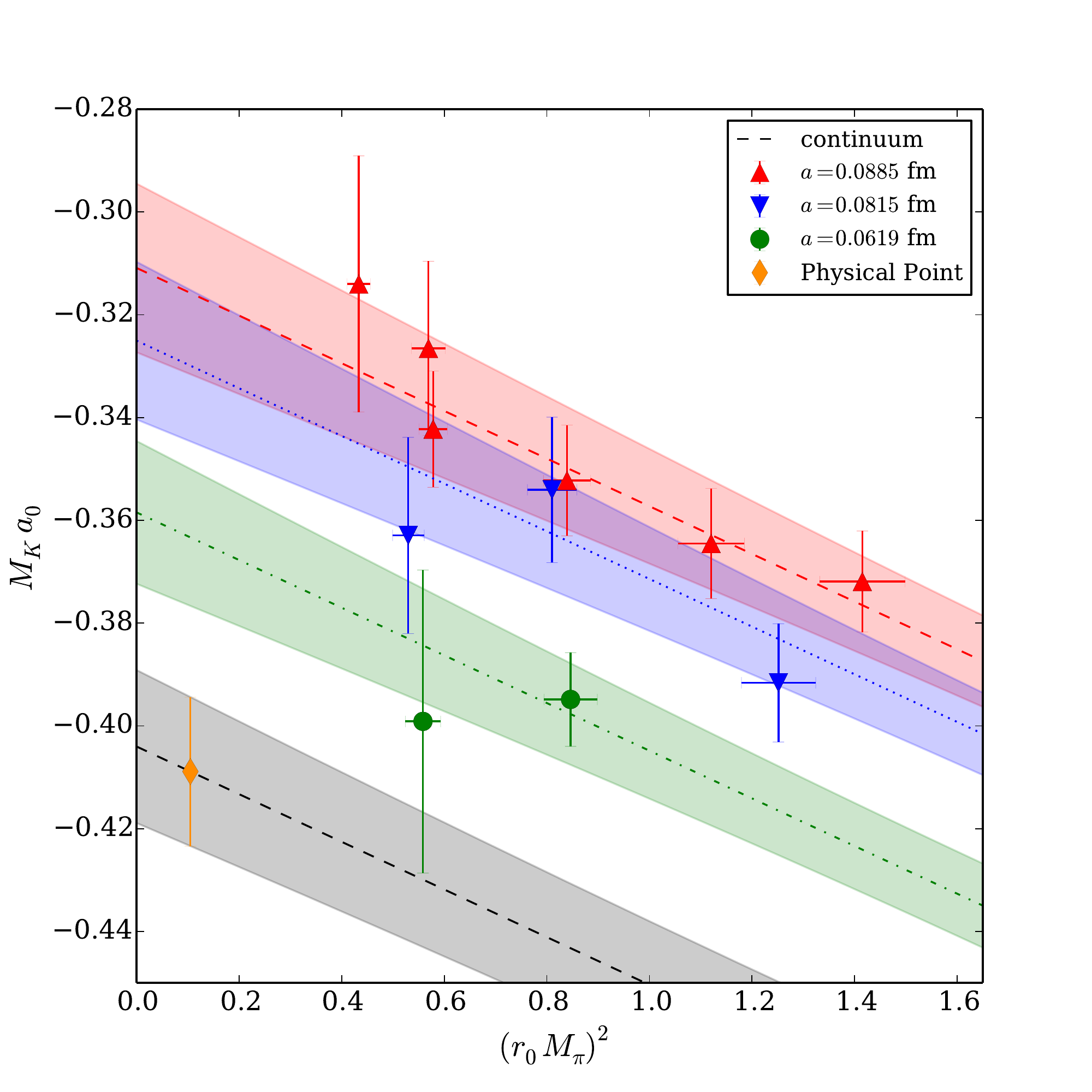}
    \caption{Chiral and continuum extrapolation of\newline $M_K a_0$ with $m_s$ fixed via
      $\Delta^2$}
    \label{fig:contA}
  \end{subfigure}
  \begin{subfigure}{.49\textwidth}
    \flushright
    \includegraphics[width=\textwidth]{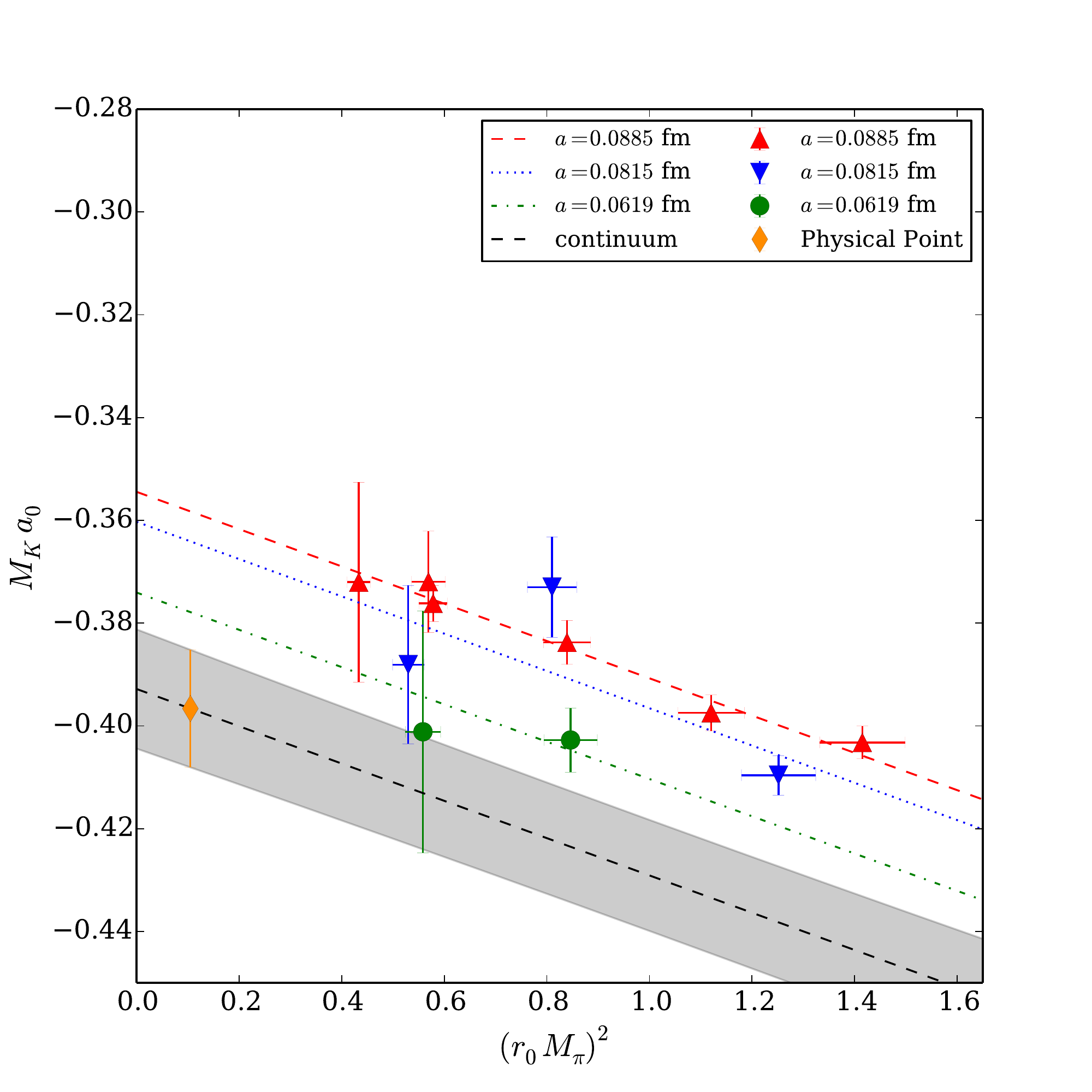}
    \caption{Same as fig.~\ref{fig:contA}, but with $m_s$ fixed via
      $M_K^{\text{exp}}$. Error bands suppressed for visibility}
    \label{fig:contB}
  \end{subfigure}
\end{figure}
\begin{figure}
\vspace{-0.5cm}
  \begin{subfigure}{.49\textwidth}
\includegraphics[width=\textwidth]{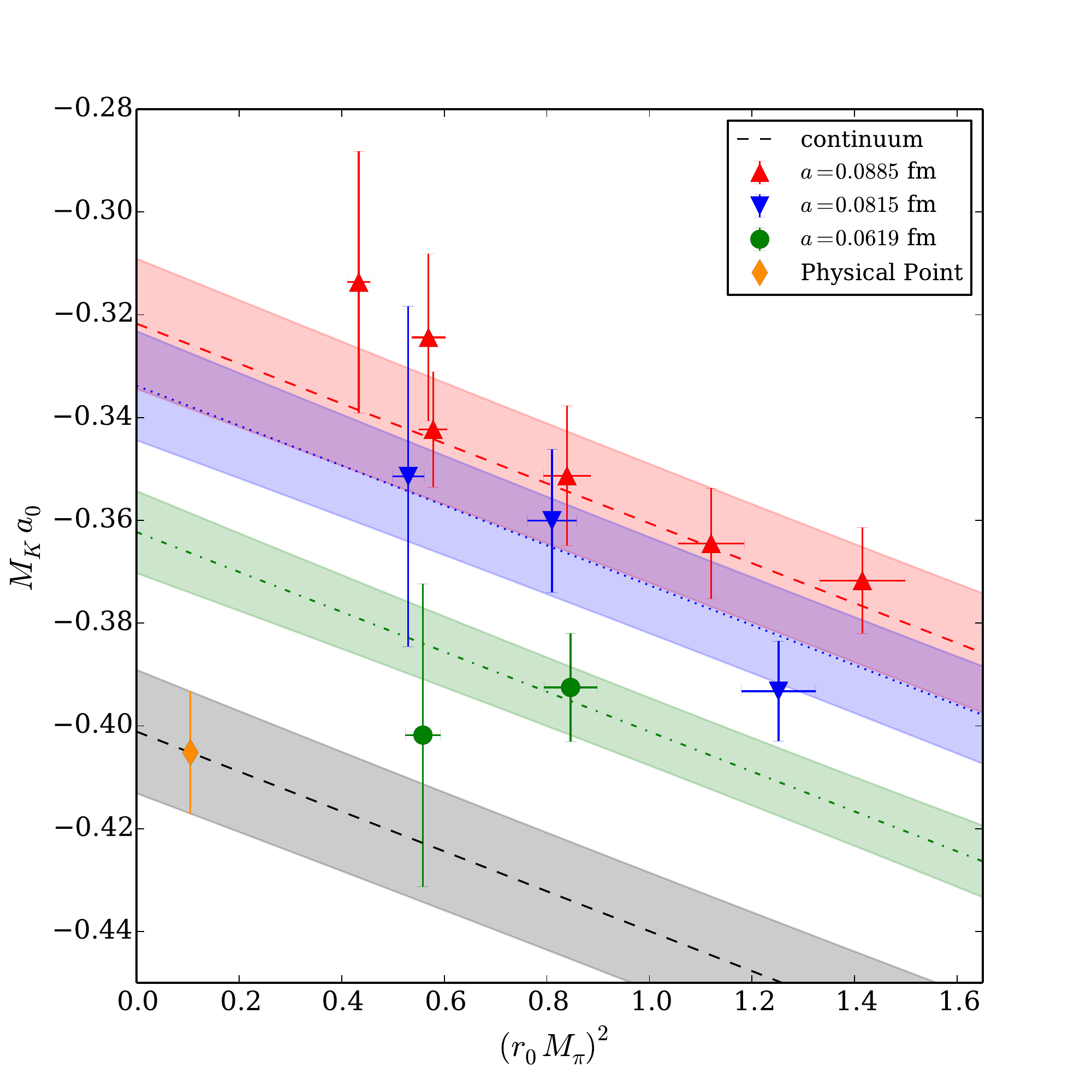}
\caption{Combined fit of eq.~\ref{eq:cont_ext} to $M_K a_0$ from method A}
\label{fig:combcontA}
  \end{subfigure}%
  \begin{subfigure}{.49\textwidth}
\includegraphics[width=\textwidth]{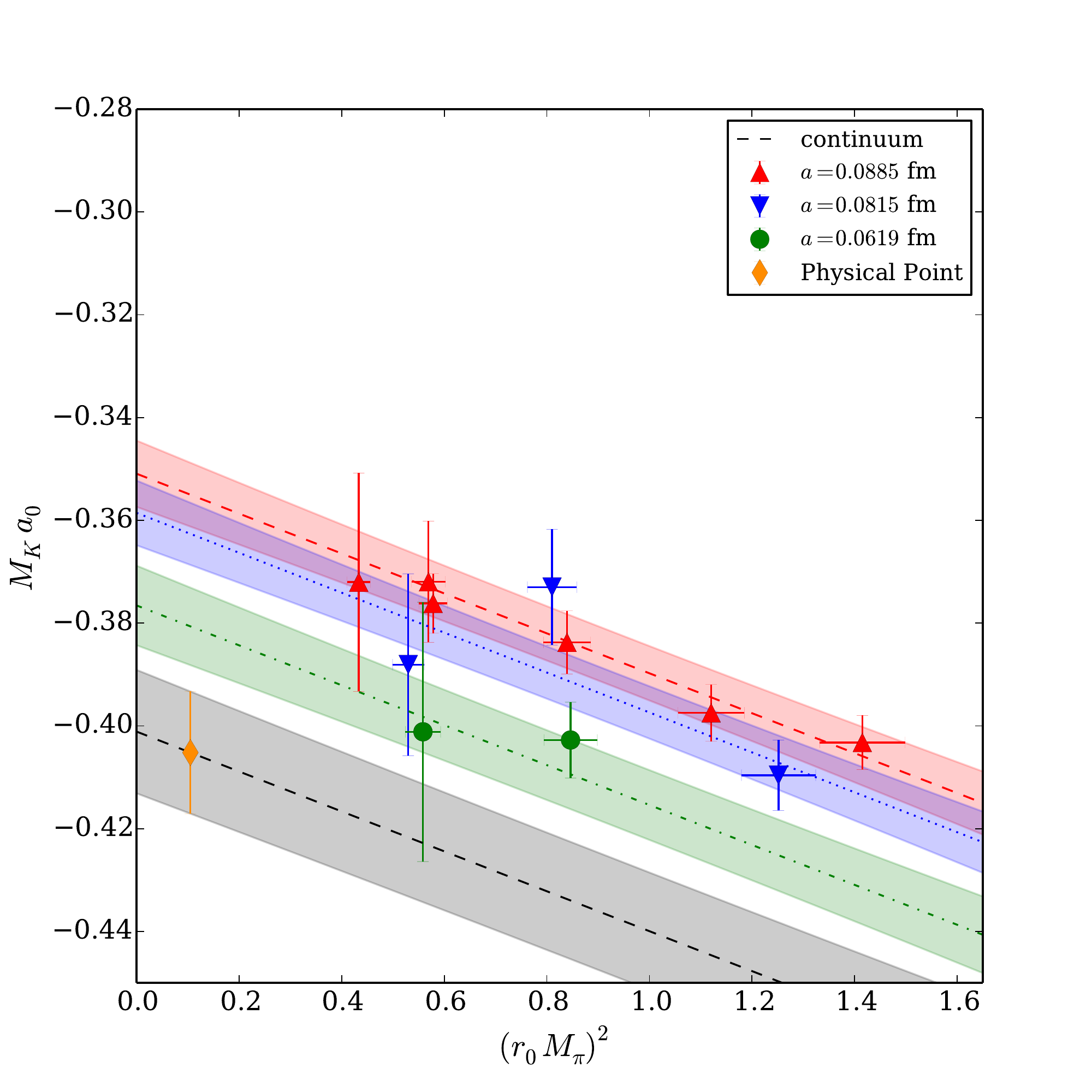}
\caption{Same as fig.~\ref{fig:combcontA} with data from method B}
\label{fig:combcontB}
  \end{subfigure}
\end{figure}

The fit results for the methods A and B are displayed in fig.~\ref{fig:contA}
and fig.~\ref{fig:contB}, respectively. We show the $m_s$-interpolated data of
$M_K a_0$ as a function of $(r_0M_{\pi}^2)$ and the lattice spacing (red, blue
and green for rising $\beta$). In addition, the fitted functions from eq.~\ref{eq:cont_ext} are shown
for each value of the lattice spacing. The continuum extrapolated
value of $M_K a_0$, extrapolated to the physical value of $M^2_\pi$ together
with the continuum version of eq.~\ref{eq:cont_ext} is displayed as well.
Despite different discretisation effects, the extrapolated values for $M_K a_0$ calculated with
method A and B agree well within errors. They are compiled in
table~\ref{tab:result1}. 
\begin{table}
  \centering
{\small
 \begin{tabular*}{.9\textwidth}{@{\extracolsep{\fill}}lcccc}
  \hline\hline
    & Method A & Method B & A$+$B combined & $p$-value weighted \\
  \hline\hline
    $(M_K a_0)_{\mathrm{phys}}$& $-0.409(15)$ & $-0.397(11)$ & $-0.405(12)$ &
    $-0.397(11)(^{+0}_{-8})$ \\
    $\chi^2$/dof & 0.38 & 0.73 & 0.65 &-\\
    $p$-value &0.93 &0.67 & 0.86 & - \\
  \hline\hline
  \end{tabular*}
}
  \caption{Comparison of the results for $M_K a_0$ at the physical point.}
  \label{tab:result1}
\end{table}
To investigate the different discretisation effects we made a combined fit of
the data from methods A and B with common parameters $P_0$ and $P_2$ and
different parameters $P_1$ and $P_1'$ for the dependence on the squared lattice
spacing $a^2$. Differences in the discretisation effects should be
taken into account by $P_1$ and $P_1'$. At the same time the parameters $P_0$ and $P_2$ do
not change. The extrapolated value of $M_K a_0$ is compatible with the ones
from method A and method B, respectively.
The outcome of the combined fit is shown in fig.~\ref{fig:combcontA}
and~\ref{fig:combcontB}. The data are the same as shown in fig.~\ref{fig:contA}
and~\ref{fig:contB}. The curves now are results from the fit of
eq.~\ref{eq:cont_ext} with the common parameters $P_0$ and $P_2$ and the
parameters $P_1$ (fig~\ref{fig:combcontA}) and $P_1'$
(fig~\ref{fig:combcontB}) coming from the different lattice spacing dependence.
The fit models both data sets equally well, confirming that the difference
stems
from discretisation effects in setting the strange quark mass. To estimate the systematic uncertainty of the Methods A
and B we calculate the $p$-value weighted median of the results from A and B.
The results are shown in tab.~\ref{tab:result1}. The 68.54\% confidence interval
of the combined and weighted distribution of A and B serves as an estimate of
the systematic uncertainty.

The NPLQCD collaboration also calculated $M_K a_0$ in ref.~\cite{Beane:2007uh}. 
The PACS-CS collaboration undertook an investigation of
scattering lengths for systems of two pseudoscalar mesons in
ref.~\cite{Sasaki:2013vxa}. 
Tab.~\ref{tab:comp} shows the weighted result for $M_K a_0$ for our work
in comparison to the calculations of NPLQCD and PACS-CS.
\begin{table}[!h]
  \centering
{\small
  \begin{tabular*}{.9\textwidth}{@{\extracolsep{\fill}}lccc}
    \hline\hline
    Collaboration & ETMC(this work) & NPLQCD & PACS-CS \\
    \hline\hline
    $M_K a_0$ & $-0.397(11)(_{-8}^{+0})$ & $-0.352(16)$ & $-0.310(10)(32)$\\
    \hline\hline
  \end{tabular*}
}
  \caption{Comparison of the results of $M_K a_0$ for this work, the work of
  NPLQCD and the work of PACS-CS. Statistical and systematic uncertainties
  shown in the first and second parentheses, respectively. For the analysis by
  NPLQCD stated uncertainty from both uncertainties added in quadrature.}
  \label{tab:comp}
\end{table}
From tab.~\ref{tab:comp} deviations beyond the statistical level of the three
analyses become visible. These differences might be explained with lattice
artifacts not taken into account by NPLQCD and PACS-CS at the same level of
control.
\section{Acknowledgements}
We would like to thank the members of the ETMC for the most enjoyable
collaboration. The computing time for this project has been made available by
the John von Neumann-Institute for Computing (NIC) on the Jureca and Juqueen
systems in J\"{u}lich. This project was funded by the DFG as a part of the
Sino-German CRC110. The open source software packages tmLQCD~\cite{Jansen:2009xp}, Boost~\cite{boost:2015} and SciPy~\cite{scipy:2001} have been used. In addition we employed QUDA~\cite{Clark:2009wm} for calculating propagators on GPUs.

\bibliographystyle{h-physrev5}
\bibliography{bibliography.bib}
\end{document}